\documentclass{aa501}

\usepackage{graphicx}
\usepackage{natbib}
\bibpunct{(}{)}{;}{a}{}{,} 

\newcommand{\Ha}{\mbox{${\rm H\alpha}$}}
\newcommand{\Hb}{\mbox{${\rm H\beta}$}}
\newcommand{\Ion}[2]{#1{\,\scriptsize #2}}
\newcommand{\Rwd}{\mbox{$R_{\rm wd}$}}
\newcommand{\Twd}{\mbox{$T_{\rm wd}$}}
\newcommand{\Porb}{\mbox{$P_{\rm orb}$}}
\newcommand{\Pspin}{\mbox{$P_{\rm spin}$}}
\newcommand{\ecs}{\mbox{$\rm erg\;cm^{-2}s^{-1}$}}
\newcommand{\gcs}{\mbox{$\rm g\;cm^{-2}s^{-1}$}}
\newcommand{\gs}{\mbox{$\rm g\;s^{-1}$}}

\newcommand{\phiorb}{\mbox{$\phi_{\rm orb}$}}
\newcommand{\phimag}{\mbox{$\phi_{\rm mag}$}}
\begin{document}

%\thesaurus{06   % A&A Section 6: Form. struct. and evolut. of stars
%	   (02.01.2;             % Accretion, accretion disks
%	    08.02.1;             % binaries close
%	    08.09.2 AM Herculis; % {\bf Stars: individual:} $\ldots$
%	    08.14.2;             % novae, cataclysmic variables
%	    08.23.1)}            % white dwarfs

\title{A model for the optical high state light curve of AM\,Herculis
\thanks{Based on observations obtained at the 1.5m Loiano telescope,
operated by the Bologna Astronomical Observatory.}}

\author{B.T. G\"ansicke\inst{1}, 
        A. Fischer\inst{1}\thanks{Present address: andreas.fischer@lhsystems.com}, 
        R. Silvotti\inst{2} \and
        D. de Martino\inst{2}}

\offprints{B. G\"ansicke, boris@uni-sw.gwdg.de}

\institute{  
Universit\"ats-Sternwarte G\"ottingen, 
Geismarlandstr. 11, D-37083 G\"ottingen, Germany
  \and    
Osservatorio di Capodimonte, Via Moiariello 16, I-80131 Napoli, Italy, demartin@na.astro.it, silvotti@na.astro.it}

\date{Received \underline{\hskip2cm} ; accepted \underline{\hskip2cm} }

\authorrunning{G\"ansicke et al.}
\titlerunning{A model for the optical high state light curve of AM\,Her}

\abstract{We present a simple quantitative model that can describe
the photometric $B$ and $V$ band light curves of AM\,Herculis obtained
during a high state. The double-humped shape of the $V$ band light
curve is dominated by cyclotron emission from a region at the main
accreting pole with an area of $\sim5\times10^{16}\mathrm{cm^2}$ and
sustaining an inflow of $\sim0.06$\,\gcs. The almost unmodulated $B$
band is dominated by emission from the accretion stream. The
contribution of the heated white dwarf to the optical emission is
small in the $B$ band, but comparable to that of the accretion stream
in the $V$ band.  The emission of the secondary star is negligible
both in $B$ and in $V$.
}
\maketitle

\section{Introduction}
In AM\,Herculis stars, or polars, a strongly magnetic white dwarf
($B\ga10$\,MG) accretes from a Roche-lobe filling late type secondary
star. The flow of matter leaving the secondary star is threaded by the
magnetic field once the magnetic pressure exceeds the ram pressure in
the accretion stream. The kinetic energy is converted into heat in a
shock near the footpoints of the flux tubes on the white dwarf
surface.  For low and intermediate {\em local} mass flow densities
($\dot m\la10\,\gcs$), this shock stands above the white
dwarf surface, and the cooling thermal bremsstrahlung and cyclotron
radiation can escape unobstacled into space. For high mass flow densities
($\dot m>10\,\gcs$), the shock is submerged into the
white dwarf atmosphere, and the primary accretion luminosity is
reprocessed into the soft X-ray regime.

The strong magnetic field of the white dwarf synchronises its rotation
with the binary orbital period ($\Pspin=\Porb$). As an observational
consequence, the emission of polars is strongly modulated at the
orbital period in almost all wavelength bands.
While the shape of the hard X-ray light curve of AM\,Herculis (and
most other polars) can be  easily interpreted in terms of the
changing geometric aspect of the hot plasma below the shock, the
optical through infrared emission may show a more complicated phase
dependence, as different components within the binary contribute to the
observed emission at a given wavelength, and as the cyclotron emission
from the accretion column is subject to wavelength-dependent beaming.

In this paper we develop a simple quantitative model which takes into
account the various emission sites in AM\,Herculis, and which can
quantitatively describe the observed $B$ and $V$ band high state light
curves with a minimum of free parameters.

\section{Observations}
We obtained $B$ and $V$ high state photometry of AM\,Her in August
1998 at the Loiano 1.5\,m telescope using a 2-channel photoelectric
photometer and a time resolution of 1\,s (Table\,\ref{t-obs}).  The
observations were interrupted every $60-120$\,min for sky measurements
in both channels. Data reduction included sky subtraction,
differential extinction, absolute calibration using Landolt standards,
and binning into 5\,s bins.
The timings have been converted to orbital phases using the ephemeris
of \citet{heise+verbunt88-1}, but defining $\phi\,(=\phiorb)=0.0$ as
the inferior conjunction of the secondary star
\citep{southwelletal95-1}\footnote{In the magnetic phase convention
which is used frequently in the literature, $\phimag=0.0$ is defined
by the linear polarization pulse observed when the angle between the
line of sight and the magnetic field line in the accretion column is
closest to $90^{\circ}$, or, alternatively, when the sign of the
circular polarization is crossing zero.}. In AM\,Her, the conversion
between the phase conventions is given by $\phiorb=\phimag+0.367$. The
high precision of the ephemeris of \citet{heise+verbunt88-1} ensures
that the error in the computed phases is
$\delta\phiorb\le2\times10^{-3}$, which is of no concern for the
purpose of the present paper.
A phenomenological description of the photometric data used here has
been included in the analysis of quasi-simultaneous BeppoSAX
observations by \citet{mattetal00-1}.

\section{Modelling the light curve}
\begin{figure}
\includegraphics[angle=270,width=8.8cm]{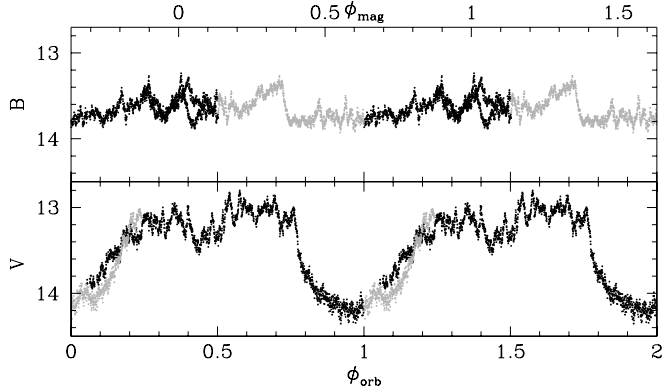}
\caption[]{\label{f-bv} Calibrated $B$ and $V$ light curves obtained
August 20 \& 22, respectively, at the Loiano 1.5\,m
telescope. Indicated are the orbital phase (bottom) and magnetic phase
(top). Both observations cover more than once the binary
orbit. The two consecutive orbits are shown by black and grey points
to highlight the cycle-to-cycle variations.}
\end{figure}

The observed light curves (Fig.\,\ref{f-bv}) are typical of AM\,Her in
the high state \citep[e.g.][]{olson77-1,szkody+brownlee77-1}.  The $V$
band shows a broad and round minimum near $\phiorb\approx0.0$ and a
secondary minimum near $\phiorb\approx0.5$. The $B$ band is almost
unmodulated. Both light curves exhibit a significant amount of
flickering on timescales of several minutes, which is probably due to
inhomogeneities in the accretion flow \citep{mattetal00-1}.

We identify four possible emission sites that will contribute to the
observed light curves: the (heated) white dwarf, the secondary
star, the accretion stream feeding material from the secondary star
towards the white dwarf, and the accretion column just above the white
dwarf surface near the magnetic pole. In the following, we will
discuss the system geometry of AM\,Her and the four individual
emission components.

\begin{table}[t]
\caption[]{Optical photometry of AM\,Her in 1998. Listed are the start
time and the the total exposure time.\label{t-obs}}
\begin{flushleft}
\begin{tabular}{lcccc}
\hline\noalign{\smallskip}
Date & Time (UT) & Exp. time [s] & Filter & Mean magnitude \\ 
\hline\noalign{\smallskip}
Aug. 20 & 20:45:21 & 13320 & $V$ & 13.50 \\
Aug. 22 & 19:45:20 & 14040 & $B$ & 13.64 \\
\noalign{\smallskip}\hline
\end{tabular}
\end{flushleft}
\end{table}

\subsection{The system geometry}
The orientation of the accretion column with respect to the observer
can be described by a set of three angles: the inclination of the
system, $i$, the co-latitude of the accretion column/spot, $\beta$, measured as
the angle between the rotation axis and the field line threading the
accretion column and the azimuth of the magnetic axis from
the line connecting the centres of the two stars, $\psi$.
For our present purpose, the exact knowledge of $\psi$ is not crucial,
as it defines only a phase offset between our model light curve and
the observed light curve, and we will adjust $\psi$ accordingly below.

The inclination and the co-latitude of AM\,Her have been
subject of dispute. \citet{brainerd+lamb85-1} derived
from the observed polarization properties $i=35^{\circ}\pm5^{\circ}$
and $\beta=58^{\circ}\pm5^{\circ}$. Another polarimetric study
indicates $i\approx50^{\circ}$ and $\beta\approx50-60^{\circ}$
\citep{wickramasingheetal91-1}. The higher value of the inclination is
favoured by spectroscopy of the secondary and of the accretion stream
\citep{davey+smith96-1,southwelletal95-1,gaensickeetal98-2}. 
An additional constraint $i+\beta\ga90^{\circ}$ comes from the
observed self-eclipse of the X-ray/EUV/ultraviolet emission from the
accretion region and from the heated surrounding white dwarf
photosphere. \citet{gaensickeetal98-2} and \citet{sirk+howell98-1}
derive independently $i+\beta=105^{\circ}$.

We will use the combinations ($i=50^{\circ}, \beta=55^{\circ}$) and
($i=35^{\circ}, \beta=70^{\circ}$) in our models of the $B$ and $V$
light curves of AM\,Her in order to reflect the uncertainties in the
exact geometry, while satisfying $i+\beta=105^{\circ}$.

\subsection{The heated white dwarf}
{\em IUE} and {\em HST} observations show that the ultraviolet flux of
AM\,Her is dominated by the emission from the accretion-heated white
dwarf during both low states and high states
\citep{gaensickeetal95-1}.  \citet{gaensickeetal98-2} could
quantitatively model the quasi-sinusoidal flux modulation observed with
{\em HST} during a high state with a large, moderately hot spot
covering a white dwarf of $\Twd=20\,000$\,K. They chose a temperature
distribution decreasing linearly in angle from the spot centre merging
into $\Twd$ at the spot radius. The best-fit spot parameters implied a
central temperature of 47\,000\,K and a spot size of $f\sim0.09$ of
the total white dwarf surface. With a distance of 90\,pc, the inferred
radius of the white dwarf was $\Rwd=1.12\times10^9$\,cm.
We use here synthetic phase-resolved spectra computed with the 3D
white dwarf model described by \citet{gaensickeetal98-2}, using their
best-fit parameters.

We note for completeness that during a \textit{low state} the $U$ and $B$
light curves of AM\,Her can very well be dominated by emission from
the heated white dwarf \citep[see e.g. Fig.\,1
of][]{bonnet-bidaudetal00-1}.

\subsection{\label{s-accretionstream}The accretion stream}
The near-ultraviolet emission ($\lambda\ga2000$\,\AA) of AM\,Her
during the high state is dominated by continuum emission from the
accretion stream. Based on phase-resolved {\em IUE} spectroscopy,
\citet{gaensickeetal95-1} showed that this near-ultraviolet emission
of the accretion stream does not vary noticeably with the orbital
phase.  In the following, we assume therefore that the accretion
stream emission does not vary over the binary orbit, and treat the
(constant) stream contributions in $B$ and $V$ as a free
parameters. The limitations implied by this assumption will be
discussed in Sect.\ref{s-discussion}.

\subsection{The secondary star} 
The secondary star in AM\,Her can be detected during low states, and
both its spectral type and magnitude are easily measured
\citep{schmidtetal81-1,gaensickeetal95-1}. For completeness,
we use here $B=18.43$ and $V=16.83$, as derived from the low state
data, to describe the contribution of the secondary. It is,
however, clear that the secondary star contribution to the observed
$B$ and $V$ \textit{high state} light curves is negligible.

\subsection{The cyclotron emission}
The accreted material releases its kinetic energy in a hydrodynamical
shock close to the white dwarf surface. The ions are heated to the
shock temperature --- a few $10^7$\,K --- while the electrons keep
at
first their pre-shock temperature because of their low mass. Coulomb
interactions between the particles transfer energy from the ions to
the electrons, such that their temperature increases below the
shock. The electrons then cool radiatively through emission of
cyclotron radiation and thermal bremsstrahlung.  This interplay
between heating and cooling results in a temperature distribution
along the post-shock flow which is characterized by its maximum
temperature \citep{woelk+beuermann96-1}.

\begin{figure}
\includegraphics[angle=270,width=8.8cm]{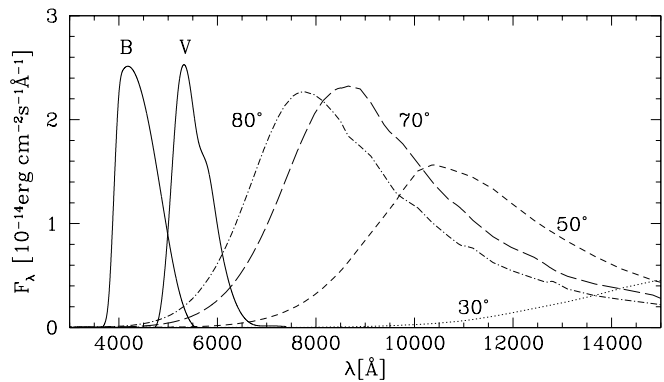}
\caption[]{\label{f-cycspec} Sample cyclotron model spectra from the
accretion column in AM\,Her for several angles $\vartheta$ between the
magnetic field line and the line-of-sight and $\dot
m=0.06$\,\gcs. Plotted are also the $B$ and $V$ response functions
(arbitarily scaled). See Fig.\,\ref{f-va} for the variation of
$\vartheta(\phi_{\rm orb})$.}
\end{figure}

\begin{figure}
\includegraphics[angle=270,width=8.8cm]{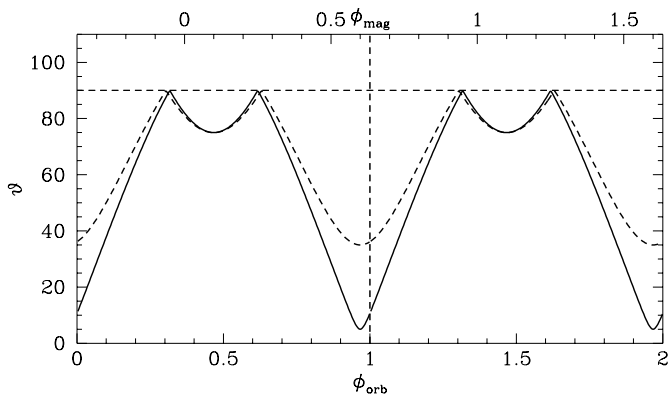}
\caption[]{\label{f-va} Angle between the magnetic field line and the
line-of-sight as function of the orbital phase. Angles $\vartheta$
exceeding $90^{\circ}$ are expressed as $180^{\circ}-\vartheta$. The
full line corresponds to $i=50^{\circ}$, $\beta=55^{\circ}$, the
dashed to $i=35^{\circ}$, $\beta=70^{\circ}$. }
\end{figure}

\begin{figure}
\centerline{\includegraphics[width=4cm]{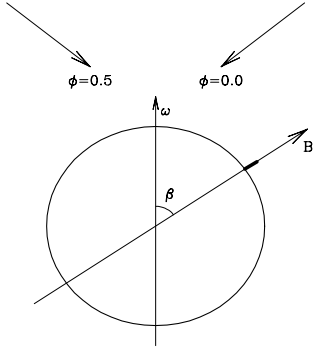}}
\caption[]{\label{f-geometry} Viewing geometry for an inclination
$i=50^{\circ}$ and a co-latitude $\beta=55^{\circ}$. $\omega$ denotes
the rotation axis, $B$ the magnetic field line in the accretion
column. The accretion column with a height of $\sim0.1$\,\Rwd\ is
plotted as bold line segment.  Indicated are the lines of sight for
inferior conjunction ($\phiorb\approx0.0$, primary minimum) and superior
conjunction ($\phiorb\approx0.5$, secondary minimum) of the secondary star.}
\end{figure}

In order to compute the emisson from the accretion column we
approximated the volume between the shock and the stellar surface with
a given cross section and height. The total emission of this column is
computed with a ray-tracing algorithm, where we solve the radiative
transfer equation for each ray passing through the structured column
using an updated version of the code described by
\citet{vanteeselingetal99-2}.  For the computation of the
angle-dependent cyclotron emission we assumed that the magnetic field
within the accretion column is perpendicular to the stellar surface
and does not vary over the considered height of the column. A
full description of the code used to solve the radiative and
hydrodynamical problem is given elsewhere \citep{fischer+beuermann01-1}.

High state X-ray observations of AM\,Her give bolometric hard X-ray
fluxes of $\sim(1-3)\times10^{-10}$\,\ecs\
\citep{gaensickeetal95-1,ishidaetal97-1,mattetal00-1}. Assuming that
the cyclotron emission can contribute roughly the same amount of flux
\citep{gaensickeetal95-1}, we estimate an accretion rate $\dot
M=10^{15}-10^{16}$\,\gs\ for the stand-off shock region\footnote{This
is \textit{not} to be confused with the total accretion rate, which
includes the regions of high local mass flow densities resulting in
submerged shocks with consecutive soft X-ray
emission. \citet{mattetal00-1} found an excess of the soft-to-hard
X-ray luminosity $L_\mathrm{SX}/L_\mathrm{HX}=5.4$.}.
The magnetic field strength of AM\,Her is $B=14\,\rm MG$
\citep{baileyetal91-1}. For the accretion rate estimated above and a
dipolar field geometry, a cross section of the accretion column of
$A_{\rm acc} \approx 5\times 10^{16}\,\rm cm^2$ is estimated near the
magnetic pole \citep{lubow+shu75-1,lubow+shu76-1,heerleinetal99-1}.
From these numbers, a mean \textit{local} mass flow density $\dot m
\approx 0.02-0.2$\,\gcs\ is derived. We have computed a grid of
spectra for several mass flow densities in the given range and for a
number of angles between the line of sight and the magnetic field. A
representative sample of spectra for different $\vartheta$ is shown in
Fig.\,\ref{f-cycspec}.

From the computed set of model spectra, we synthesized $B$ and $V$
light curves of the cyclotron emission from the accretion column,
where the mass flow density $\dot m$ was taken as a free parameter.
The fundamental parameter in the computation of the observed cyclotron
flux for a given orbital phase is the angle between the line of sight
and the magnetic field axis, $\vartheta$. For a given set of
inclination $i$, co-latitutde $\beta$ and orbital phase $\phiorb$
\begin{equation}
\cos \vartheta = \cos i \cos\beta - \sin i \sin\beta\cos(\phiorb+\pi/2-\psi)
\end{equation}

Figure\,\ref{f-va} shows the variation of $\vartheta$ with $\phiorb$
for ($i=50^{\circ}, \beta=55^{\circ}$) and for ($i=35^{\circ},
\beta=70^{\circ}$), the viewing geometry is illustrated in
Fig.\,\ref{f-geometry}. We chose $\psi=-12^{\circ}$ so that the
minimum in $\vartheta$ aligns with the minimum of the observed $V$
light curve, i.e. the magnetic pole leads the secondary by
$\psi$. This value is in good agreement with the azimuth of the hot
accretion region derived from X-ray and ultraviolet light curves,
$\psi_\mathrm{HS}\approx-10^{\circ}\mathrm{~to~}0^{\circ}$ 
\citep{paerelsetal96-1,gaensickeetal98-2}.
$\vartheta$ reaches a minimum near $\phiorb\approx-0.03$, when the
observer looks down along the accretion funnel, passes through
$90^{\circ}$ at $\phiorb\approx0.3$ and reaches the maximum of
$105^{\circ}$ at $\phiorb\approx0.47$. As the cyclotron beaming is
sensitive only to $|\cos\vartheta|$ we plot $180^{\circ}-\vartheta$
when $\vartheta>90^{\circ}$. From Fig.\,\ref{f-va} it is clear that a
{\em reduction} of the cyclotron flux is expected in the range
$\phiorb\approx0.3-0.6$ from cyclotron beaming alone, without taking
into account a possible occultation of the accretion column by the
body of the white dwarf (see also Sect.\,\ref{s-discussion}).

A very similar \textit{qualitative} description of the phase-dependent
cyclotron emission has been given by \citet{baileyetal84-1} in order
to explain the polarimetric properties of AM\,Her: linear polarization
pulses and a change of sign of the circular polarization are observed
at $\phimag\approx0.0$ and $0.3$, which corresponds to the phases when
$\vartheta\simeq90^{\circ}$.

\subsection{Composite model light curves\label{s-lightcurvemodel}}
\begin{figure}
\includegraphics[angle=270,width=8.8cm]{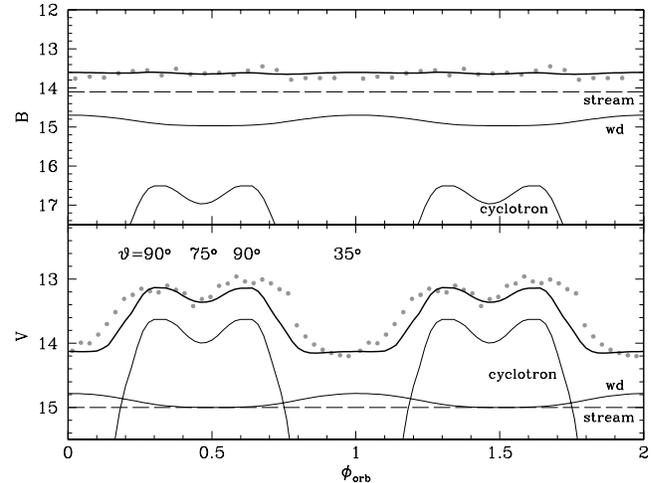}
\caption[]{\label{f-fit1} The observed light curves from
Fig.\,\ref{f-bv} have been sampled into $\Delta\phiorb=0.03$ orbital
phase bins (grey points). The composite model light curves assuming
$i=50^{\circ}$, $\beta=55^{\circ}$ are shown as fat black lines.
Contributions to the total fluxes come from the heated white dwarf,
from the accretion column, and from the accretion stream (the
secondary star contributes too little to show up in the graph).  The
angles $\vartheta$ indicate the angle between the magnetic field and
the line-of-sight (cf. Fig\,\ref{f-va}).}
\end{figure}

\begin{figure}
\includegraphics[angle=270,width=8.8cm]{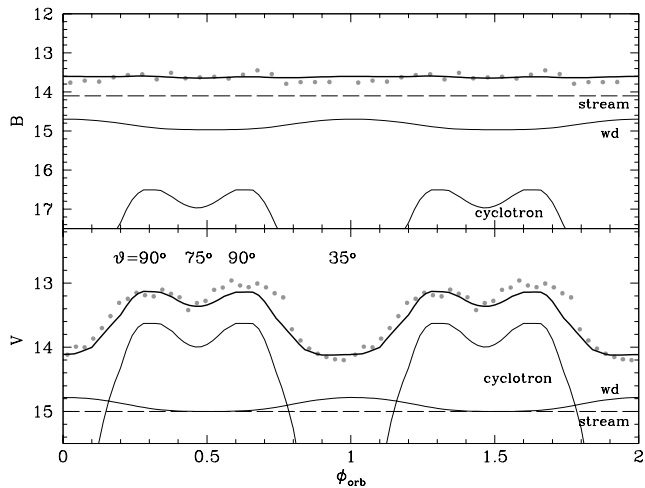}
\caption[]{\label{f-fit2} As Fig.\,\ref{f-fit1}, but assuming $i=35^{\circ}$,
$\beta=70^{\circ}$.}
\end{figure}

We obtain the model light curves by summing up the  four
individual emission components, white dwarf, accretion stream,
secondary star, and accretion column. 
To fit the flux and the shape of the observed $B$ and $V$ light
curves, we use two free parameters: the mean local mass flow density
$\dot m$ which determines the (phase-dependent) contribution of the
cyclotron emission and the (phase-independent) contribution of the
accretion stream, $B_\mathrm{stream}$ and $V_\mathrm{stream}$.  As
mentioned above, the contributions of the heated white dwarf and of
the secondary star are fixed. The observed light curves are reasonably
well reproduced for $\dot m=0.06$\,\gcs, $B_\mathrm{stream}=15.0$, and
$V_\mathrm{stream}=14.1$ (Fig.\,\ref{f-fit1},\ref{f-fit2}).
Figures\,\ref{f-fit1} and \ref{f-fit2} show the model light curves for
the two different choices of ($i, \beta$). The contribution of the
heated white dwarf to the observed $B$ and $V$ high state light curves
is small (Fig.\,\ref{f-fit1},\ref{f-fit2}), the contribution of the
secondary is negligible.

Both fits grossly reproduce the shape of the observed light curves,
i.e. a broad and deep minimum at $\phiorb=0.0$ and a shallow secondary
minimum at $\phiorb=0.5$ in $V$ and practically no orbital modulation
in $B$. However, for ($i=50^{\circ}, \beta=55^{\circ}$) the angle
$\vartheta$ drops below $40^{\circ}$ during $\phi=0.8-1.1$, with the
result that cyclotron radiation contributes very little to the
total $V$ band emission. Consequently, this model light curve has a
flat bottom, which contrasts the observed round shape.

\section{Discussion\label{s-discussion}}
We have made a number of simplifications and assumptions in our light
curve model that need some further attention.

(1) We have assumed that the accretion stream emission does not vary
over the orbit, and have treated its contribution as a free parameter
to fit the observed flux level. The computation of realistic accretion
stream spectra is a problem of great complexity
\citep[e.g.][]{ferrario+wehrse99-1}, and certainly beyond the scope of
our paper. However, a short inspection of the general spectral
features of accretion stream emission lends support to our assumption.

The optical high state spectrum of AM\,Her is characterized by strong
Balmer and Helium emission lines and by a strong Balmer jump in
emission \citep[e.g.][]{schachteretal91-1}. The origin of such a
spectrum is clearly a plasma in which the optical depth in the continuum
is below unity (\textit{optically thin} in the continuum), and the
optical depth in the lines is much larger than in the continuum,
i.e. $\gg1$ (\textit{optically thick} in the lines). The details of
the spectrum depend on the temperature, the density, and the extension
of the plasma region, but the general characteristics are (a)
significant continuum emission in the near-ultraviolet part of the
Balmer continuum, (b) a strong contribution in the $B$ band, which
covers the higher lines of the Balmer series, \Ion{Ca}{K+H} in
emission, as well as the Balmer jump, and (c) a weak contribution in
the $V$ band, which covers practically only the weak continuum between
\Ha\ and \Hb.
As already stated in Sect.\,\ref{s-accretionstream},
\citet{gaensickeetal95-1} detected strong near-ultraviolet continuum
emission with a weak phase-dependence, and identified the accretion
stream as the source of this emission. In AM\,Her, with
$B\approx14$\,MG, cyclotron emission can not contribute significantly
at such short wavelengths, the secondary star can be excluded, and the
white dwarf contribution was already modelled in the analysis of
\citet{gaensickeetal95-1}.  We feel, therefore, safe to identify the
observed near-ultraviolet continuum flux with the accretion stream
feature (a) defined above.

We find in our present analysis that the accretion stream contributes
more in $B$ than in $V$. In fact, the stream emission dominates the
practically flat $B$ band light curve. This result appears entirely
plausible, as our above considerations show that the stream is
indeed expected to be brigher in $B$ than in $V$. We note that weak
photometric modulation at blue wavelengths is a hallmark of AM\,Her
during the high state \citep[e.g.][]{olson77-1}.
 
A more sophisticated approach would need simultaneous phase-resolved
spectroscopy, so that the stream contribution in the $B$ and $V$ bands
could be quantitatively estimated e.g.  from the orbital variation of
the emission line fluxes. Hence, in summary, the observed weak phase
dependence of the near-ultraviolet stream emission and the fact that
stream is expected to be brighter in $B$ than in $V$ strongly supports
our assumption of a phase-independent stream contribution.  This in
turn implies, as shown in Sect.\,\ref{s-lightcurvemodel}, that
cyclotron emission is the dominant \textit{phase-dependent}
contribution to the observed $V$ band high state light curve.

(2) We ignore the geometric occultation of the accretion column by the
body of the white dwarf. For $i+\beta=105^{\circ}$ structures
extending less than than $\sim0.035$\,\Rwd\ above the white dwarf
surface will be totally eclipsed at $\phiorb\approx0.5$. However, for
the low magnetic field and the rather low mass flow density, the shock
is expected to stand $\sim0.1$\,\Rwd\ above the white dwarf surface
\citep{fischer+beuermann01-1}. The X-ray data of AM\,Her
clearly show that at least part of the accretion region \textit{is}
eclipsed at $\phiorb\approx0.5$ (i.e. when $\vartheta$ reaches its
maximum) by the white dwarf \cite[e.g.][]{hearn+richardson77-1}. In
contrast to this, significant polarized emission is still detected
during this phase, indicating either that the cyclotron emission
arises at a greater height above the white dwarf surface than the
X-rays \citep{baileyetal84-1}, or that additional cyclotron emission
comes from the opposite magnetic pole \citep{wickramasingheetal91-1}.
An indication that an eclipse of the accretion column by the white
dwarf might be relevant comes from the $UBVRI$ polarimetry of
\citet{piirolaetal85-1}: the secondary minimum at $\phiorb\approx0.6$
has practically the same shape in $V$, $R$, and $I$. This is
unexpected, as the cyclotron beaming decreases towards the lower
cyclotron harmonics (i.e. longer wavelengths). In fact, $R$ and $I$ band
light curves calculated with our model give reasonable magnitudes, but
do not show a strong secondary minimum.

(3) We have neglected that the accretion column has a finite lateral
extent, which will cause a variation of the magnetic field strength
and, more importantly, of its  orientation within the cyclotron emitting
region. The expected result of such variation is a smoothing of
$\vartheta(\phiorb)$, and, consequently, of the cyclotron light curve.
In addition, a significant lateral extent of the cyclotron
emitting region would counteract to some extent the self-eclipse
just described above.

(4) We deliberately restrict our model to a single cyclotron region
near the main (upper) pole. The X-ray emission of AM\,Her is known to
switch between (at least, see below) two different patterns. In the
normal mode hard and soft X-rays are emitted in phase from the upper
pole. During the reversed mode, the soft X-ray emission originates
predominantly at the lower pole, while the hard X-rays are still
emitted from the upper pole \citep{heiseetal85-1}.
Despite this apparently drastic change in the accretion pattern, the
optical light curve of AM\,Her does \textit{not} change between normal
and reversed mode \citep{mazehetal86-1}. 
These observations strongly suggest that only the high mass flow
densities may switch back and forth between the poles, while the
upper pole is fed during both states by low mass flow densities
--~associated with the emission of hard X-ray and cyclotron
emission. This interpretation supports our one-pole cyclotron emission
model.

An ultimate puzzle, though, remains: while our optical data from 1998,
August 20/22 presented here are completely typical of AM\,Her in
either the normal mode or the reversed mode, they were obtained only
about a week after our BeppoSAX observations (1998, August 12) which
showed the system in an hitherto unknown state with significant hard
X-ray emission during $\phiorb\simeq0.5$ \citep{mattetal00-1}. RXTE
and EUVE observations obtained on 1998, August 4 confirmed the
atypical X-ray light curve \citep{christian00-1}. We are left with the
possibility that AM\,Her returned into one of the previously known
--~normal or reversed~-- modes by the time of our optical campaign.
Considering that also the optical high state light curve of AM\,Her
noticeably changed on a (to our knowledge) single occasion
\citep{szkody78-1}, it appears that AM\,Her occasionally drops into
short-lived ``atypical'' behaviour. A multiwavelength campaing
covering one of these  ``atypical'' states could reveal whether they are
related to the switching between the normal and the reversed accretion
mode.

\acknowledgements{We thank Klaus Beuermann for many discussions on
this topic and the referee Steve Howell for a number of helpful
comments.  This research was supported by the DLR under grant
50\,OR\,99\,03\,6.}

\bibliographystyle{apj}
%\bibliography{aamnem99,/home/raid1/boris/cthulhu/tex/Papers/Bibliography/aabib}

\begin{thebibliography}{31}
\expandafter\ifx\csname natexlab\endcsname\relax\def\natexlab#1{#1}\fi

\bibitem[{{Bailey} {et~al.}(1991){Bailey}, {Ferrario}, \&
  {Wickramasinghe}}]{baileyetal91-1}
{Bailey}, J., {Ferrario}, L., \& {Wickramasinghe}, D.~T. 1991, MNRAS, 251, 37P

\bibitem[{{Bailey} {et~al.}(1984){Bailey}, {Hough}, {Gilmozzi}, \&
  {Axon}}]{baileyetal84-1}
{Bailey}, J., {Hough}, J.~H., {Gilmozzi}, R., \& {Axon}, D.~J. 1984, MNRAS,
  207, 777

\bibitem[{{Bonnet-Bidaud} {et~al.}(2000){Bonnet-Bidaud}, {Mouchet},
  {Shakhovskoy}, {Somova}, {Somov}, {Andronov}, {de Martino}, {Kolesnikov}, \&
  {Kraicheva}}]{bonnet-bidaudetal00-1}
{Bonnet-Bidaud}, J.~M., {Mouchet}, M., {Shakhovskoy}, N.~M., {Somova}, T.~A.,
  {Somov}, N.~N., {Andronov}, I.~L., {de Martino}, D., {Kolesnikov}, S.~V., \&
  {Kraicheva}, Z. 2000, A\&A, 354, 1003

\bibitem[{{Brainerd} \& {Lamb}(1985)}]{brainerd+lamb85-1}
{Brainerd}, J.~J. \& {Lamb}, D.~Q. 1985, in Cataclysmic Variables and Low-Mass
  X-ray Binaries, ed. D.~Q. {Lamb} \& J.~{Patterson} (Dordrecht: D. Reidel),
  247--256

\bibitem[{{Christian}(2000)}]{christian00-1}
{Christian}, D.~J. 2000, AJ, 119, 1930

\bibitem[{{Davey} \& {Smith}(1996)}]{davey+smith96-1}
{Davey}, S.~C. \& {Smith}, R.~C. 1996, MNRAS, 280, 481

\bibitem[{{Ferrario} \& {Wehrse}(1999)}]{ferrario+wehrse99-1}
{Ferrario}, L. \& {Wehrse}, R. 1999, MNRAS, 310, 189

\bibitem[{{Fischer} \& {Beuermann}(2001)}]{fischer+beuermann01-1}
{Fischer}, A. \& {Beuermann}, K. 2001, A\&A, {\em submitted}

\bibitem[{{G\"ansicke} {et~al.}(1995){G\"ansicke}, {Beuermann}, \& {de
  Martino}}]{gaensickeetal95-1}
{G\"ansicke}, B.~T., {Beuermann}, K., \& {de Martino}, D. 1995, A\&A, 303, 127

\bibitem[{{G\"ansicke} {et~al.}(1998){G\"ansicke}, {Hoard}, {Beuermann},
  {Sion}, \& {Szkody}}]{gaensickeetal98-2}
{G\"ansicke}, B.~T., {Hoard}, D.~W., {Beuermann}, K., {Sion}, E.~M., \&
  {Szkody}, P. 1998, A\&A, 338, 933

\bibitem[{{Hearn} \& {Richardson}(1977)}]{hearn+richardson77-1}
{Hearn}, D.~R. \& {Richardson}, J.~A. 1977, ApJ Lett., 213, L115

\bibitem[{{Heerlein} {et~al.}(1999){Heerlein}, {Horne}, \&
  {Schwope}}]{heerleinetal99-1}
{Heerlein}, C., {Horne}, K., \& {Schwope}, A.~D. 1999, MNRAS, 304, 145

\bibitem[{{Heise} {et~al.}(1985){Heise}, {Brinkman}, {Gronenschild}, {Watson},
  {King}, {Stella}, \& {Kieboom}}]{heiseetal85-1}
{Heise}, J., {Brinkman}, A.~C., {Gronenschild}, E., {Watson}, M., {King},
  A.~R., {Stella}, L., \& {Kieboom}, K. 1985, A\&A, 148, L14

\bibitem[{{Heise} \& {Verbunt}(1988)}]{heise+verbunt88-1}
{Heise}, J. \& {Verbunt}, F. 1988, A\&A, 189, 112

\bibitem[{{Ishida} {et~al.}(1997){Ishida}, {Matsuzaki}, {Fujimoto}, {Mukai}, \&
  {Osborne}}]{ishidaetal97-1}
{Ishida}, M., {Matsuzaki}, K., {Fujimoto}, R., {Mukai}, K., \& {Osborne}, J.~P.
  1997, MNRAS, 287, 651

\bibitem[{{Lubow} \& {Shu}(1975)}]{lubow+shu75-1}
{Lubow}, S.~H. \& {Shu}, F.~H. 1975, ApJ, 198, 383

\bibitem[{{Lubow} \& {Shu}(1976)}]{lubow+shu76-1}
---. 1976, ApJ Lett., 207, L53

\bibitem[{{Matt} {et~al.}(2000){Matt}, {de Martino}, {G\"ansicke},
  {Negueruela}, {Silvotti}, {Bonnet-Bidaud}, {Mouchet}, \&
  {Mukai}}]{mattetal00-1}
{Matt}, G., {de Martino}, D., {G\"ansicke}, B.~T., {Negueruela}, I.,
  {Silvotti}, R., {Bonnet-Bidaud}, J.~M., {Mouchet}, M., \& {Mukai}, K. 2000,
  A\&A, 358, 177

\bibitem[{{Mazeh} {et~al.}(1986){Mazeh}, {Kieboom}, \& {Heise}}]{mazehetal86-1}
{Mazeh}, T., {Kieboom}, K., \& {Heise}, J. 1986, MNRAS, 221, 513

\bibitem[{{Olson}(1977)}]{olson77-1}
{Olson}, E.~C. 1977, ApJ, 215, 166

\bibitem[{{Paerels} {et~al.}(1996){Paerels}, {Hur}, {Mauche}, \&
  {Heise}}]{paerelsetal96-1}
{Paerels}, F., {Hur}, M.~Y., {Mauche}, C.~W., \& {Heise}, J. 1996, ApJ, 464,
  884

\bibitem[{{Piirola} {et~al.}(1985){Piirola}, {Vilhu}, {Kyrolainen},
  {Shakhovskoi}, \& {Efimov}}]{piirolaetal85-1}
{Piirola}, V., {Vilhu}, O., {Kyrolainen}, J., {Shakhovskoi}, N., \& {Efimov},
  Y. 1985, in Recent Results on Cataclysmic Variables, ESA--SP No. 236
  (Noordwijk: ESA Publications Division), 245--249

\bibitem[{{Schachter} {et~al.}(1991){Schachter}, {Filippenko}, {Kahn}, \&
  {Paerels}}]{schachteretal91-1}
{Schachter}, J., {Filippenko}, A.~V., {Kahn}, S.~M., \& {Paerels}, F. B.~S.
  1991, ApJ, 373, 633

\bibitem[{{Schmidt} {et~al.}(1981){Schmidt}, {Stockman}, \&
  {Margon}}]{schmidtetal81-1}
{Schmidt}, G.~D., {Stockman}, H.~S., \& {Margon}, B. 1981, ApJ Lett., 243, L157

\bibitem[{{Sirk} \& {Howell}(1998)}]{sirk+howell98-1}
{Sirk}, M.~M. \& {Howell}, S.~B. 1998, ApJ, 506, 824

\bibitem[{{Southwell} {et~al.}(1995){Southwell}, {Still}, {Connon Smith}, \&
  {Martin}}]{southwelletal95-1}
{Southwell}, K.~A., {Still}, M.~D., {Connon Smith}, R., \& {Martin}, J.~S.
  1995, A\&A, 302, 90

\bibitem[{{Szkody}(1978)}]{szkody78-1}
{Szkody}, P. 1978, PASP, 90, 61

\bibitem[{{Szkody} \& {Brownlee}(1977)}]{szkody+brownlee77-1}
{Szkody}, P. \& {Brownlee}, D.~E. 1977, ApJ Lett., 212, L113

\bibitem[{{van Teeseling} {et~al.}(1999){van Teeseling}, {Fischer}, \&
  {Beuermann}}]{vanteeselingetal99-2}
{van Teeseling}, A., {Fischer}, A., \& {Beuermann}, K. 1999, in Annapolis
  Workshop on Magnetic Cataclysmic Variables, ed. C.~{Hellier} \& K.~{Mukai}
  (ASP Conf. Ser. 157), 309--316

\bibitem[{{Wickramasinghe} {et~al.}(1991){Wickramasinghe}, {Bailey}, {Meggitt},
  {Ferrario}, {Hough}, \& {Tuohy}}]{wickramasingheetal91-1}
{Wickramasinghe}, D.~T., {Bailey}, J., {Meggitt}, S. M.~A., {Ferrario}, L.,
  {Hough}, J., \& {Tuohy}, I.~R. 1991, MNRAS, 251, 28

\bibitem[{{Woelk} \& {Beuermann}(1996)}]{woelk+beuermann96-1}
{Woelk}, U. \& {Beuermann}, K. 1996, A\&A, 306, 232

\end{thebibliography}

\end{document}